\documentclass[aps, prl, twocolumn, superscriptaddress, english]{revtex4}
\usepackage{graphicx}
\usepackage{cancel}
\usepackage{enumerate}
\usepackage{hyperref}
\usepackage{textcomp}
\usepackage{amsmath}
\usepackage{amssymb}
\usepackage{pgf}
\usepackage{soul}
\usepackage{subfigure}
\usepackage[english]{babel}

\newcommand{\bbeta}{\boldsymbol{\beta}}
\newcommand{\bgamma}{\boldsymbol{\gamma}}

\def\ket#1{| #1\rangle}

\newcommand{\nep}{\textrm{e}}

\newcommand{\target}{\mathrm{\scriptscriptstyle targ}}

\newcommand{\PauliSigma}{\hat{\sigma}}

\newcommand{\Ho}{\hat{H}}

\newcommand{\Ham}{\widehat{H}}
\newcommand{\Hz}{\widehat{H}_z}
\newcommand{\Hx}{\widehat{H}_x}

\newcommand{\Ptrot}{\mathrm{P}}

\begin{document}

\title{Reinforcement Learning assisted Quantum Optimization}
\author{Matteo M. Wauters}
\affiliation{SISSA, Via Bonomea 265, I-34136 Trieste, Italy}
\author{Emanuele Panizon}
\affiliation{Fachbereich Physik, Universit\"at Konstanz, 78464 Konstanz, Germany}
\author{Glen B. Mbeng}
\affiliation{Universit\"at Innsbruck, Technikerstra{\ss}e 21 a, A-6020 Innsbruck, Austria}
\author{Giuseppe E. Santoro}
\affiliation{SISSA, Via Bonomea 265, I-34136 Trieste, Italy}
\affiliation{International Centre for Theoretical Physics (ICTP), P.O.Box 586, I-34014 Trieste, Italy}
\affiliation{CNR-IOM Democritos National Simulation Center, Via Bonomea 265, I-34136 Trieste, Italy}

\begin{abstract}
We propose a reinforcement learning (RL) scheme for feedback quantum control within the quantum approximate optimization algorithm (QAOA).
QAOA requires a variational minimization for states constructed by applying a sequence of unitary operators, depending on parameters 
living in a highly dimensional space.
We reformulate such a minimum search as a learning task, where a RL agent chooses the control parameters for the unitaries, 
given partial information on the system.
We show that our RL scheme finds a policy converging to the optimal adiabatic solution for QAOA found by Mbeng {\em et al.} arXiv:1906.08948
for the translationally invariant quantum Ising chain.  
In presence of disorder, we show that our RL scheme allows the training part to be performed on small samples, and transferred successfully on larger systems. 
%
\end{abstract}

\maketitle

{\em Introduction --- }
Quantum optimization and control are at the leading edge of current research in quantum computation~\cite{Nielsen_Chuang:book}. 
Quantum Annealing (QA)~\cite{Finnila_CPL94, Kadowaki_PRE98, Brooke_SCI99, Santoro_SCI02, Santoro_JPA06}, 
{\em alias} Adiabatic Quantum Computation (AQC) \cite{Farhi_SCI01, Albash_RevModPhys2018}, is a promising quantum algorithm implemented~\cite{Johnson_Nat11}
in present noisy intermediate-scale quantum devices~\cite{Preskill_Quantum2018}. 
More recently, the Quantum Approximate Optimization Algorithm (QAOA)~\cite{Farhi_arXiv2014} --- a hybrid quantum-classical variational optimization scheme~\cite{McClean_NewJPhys2016} --- has gained momentum ~\cite{Lloyd_arXiv2018,Zhou_arXiv2018,Mbeng_arx19,morales_arxiv2019} 
and has been successfully realized in several experimental platforms~\cite{Pagano_arXiv2019,arute2020quantum}.

In QA/AQC one constructs an interpolating Hamiltonian $\Ham(s) = s \Ho_z + (1-s) \Ho_x$, where, {\em e.g.}, for spin-1/2 systems $\Ho_z$ is the problem Hamiltonian 
whose ground state (GS) we are searching~\cite{Lucas2014} while $\Ho_x = - h \sum_j \PauliSigma^x_j$  is a transverse field term. 
An adiabatic dynamics is then attempted by slowly increasing $s(t)$ from $s(0)=0$ to $s(\tau)=1$ in a large annealing time $\tau$, 
starting from some easy-to-prepare initial state $|+\rangle$, the GS of $\Ho_x$. 
The difficulty is usually associated with the growing annealing time $\tau$ necessary when the system crosses a transition point, especially of first order~\cite{Zamponi_QA:review}.

QAOA, instead, uses a variational {\em Ansatz} of the form  
%
%
\begin{equation} \label{eq:QAOA1}
|\psi_{\Ptrot}(\bgamma,\bbeta )\rangle 
= \Big( \prod_{t}^{\Ptrot \leftarrow 1} \nep^{-i \beta_{t} \Hx } \nep^{-i \gamma_{t} \Hz} \Big) |+\rangle \;,
\end{equation}
where $\bgamma=\gamma_1,\dots,\gamma_{\Ptrot}$ and $\bbeta=\beta_1,\dots,\beta_{\Ptrot}$ are $2\Ptrot$ real parameters. 
The variational state $|\psi_{\Ptrot}(\bgamma,\bbeta )\rangle$ is as a sequence of quantum gates, corresponding to 2$\Ptrot$ unitary 
evolution operators applied to the initial state, from right to left for increasing $t=1,\dots,\Ptrot$, each parameterized by control parameters 
$\gamma_{t}$ or $\beta_{t}$.
The standard QAOA approach consists of a classical minimum search in such a $2\Ptrot$-dimensional energy landscape, 
which is in general not a trivial task~\cite{McClean_NatCom2018}.
Indeed, there are in general very many local minima in the QAOA-landscape, and local optimizations with random starting points produce irregular parameter 
sets $\left( \bgamma^*,\bbeta^* \right)$, hard to implement and sensitive to noise.
To obtain stable and regular solutions $\left( \bgamma^*,\bbeta^* \right)$ that can be easily generalized to different values of $\Ptrot$ and implemented experimentally, 
it is necessary to employ iterative procedures during the minimum search~\cite{Zhou_arXiv2018,Pagano_arXiv2019,Mbeng_arx19}.
Interestingly, as discovered in Ref.~\cite{Mbeng_arx19} for quantum Ising chains, smooth regular optimal schedules for $\gamma_t$ and $\beta_t$ can be found,
which are {\em adiabatic} in a digitized-QA/AQC \cite{Martinis_Nat16} context.   

\begin{figure}[h]
\begin{center}
\includegraphics[width=8cm]{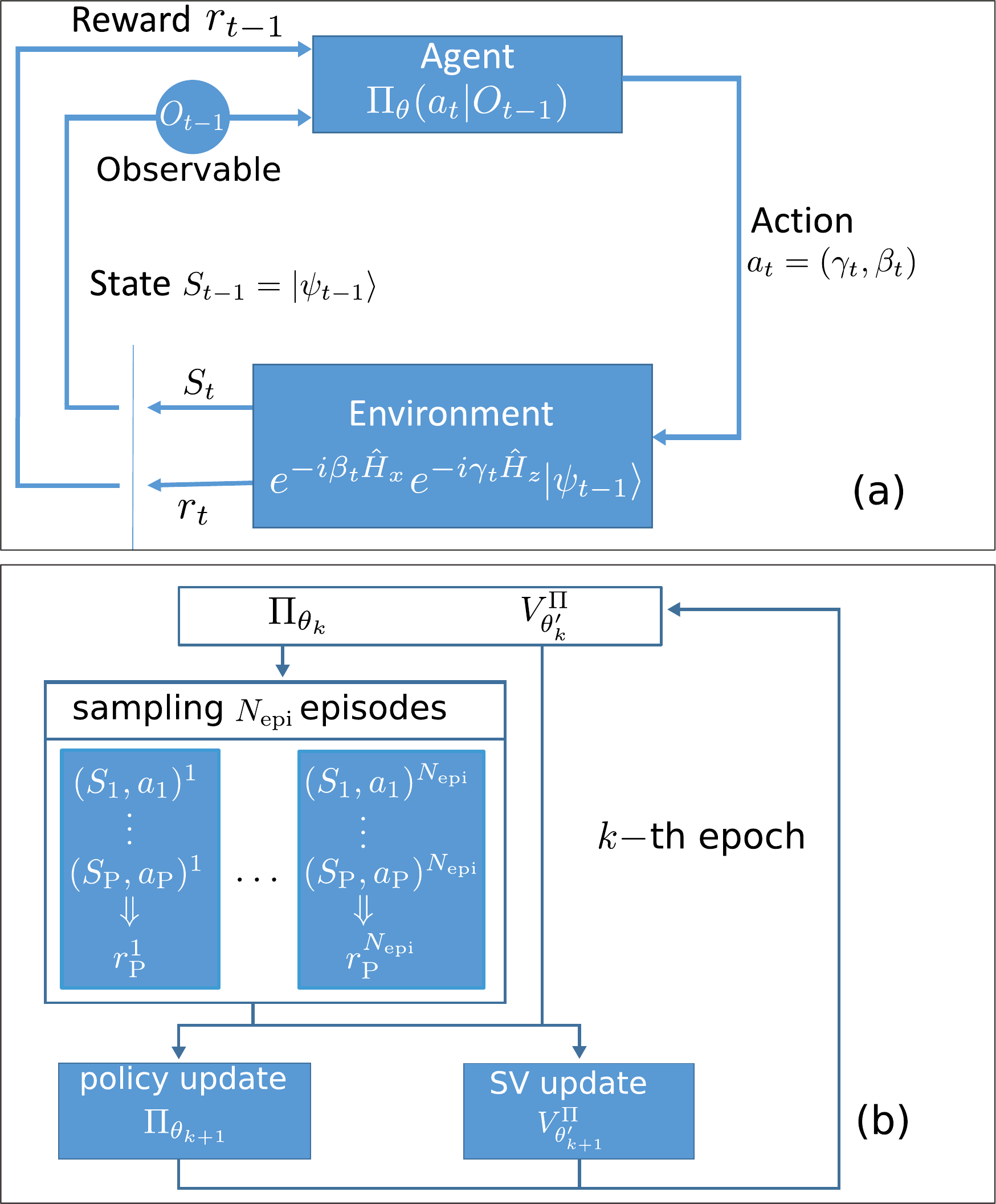}
\caption{Scheme of: (a) a single step of Reinforcement Learning for QAOA;
(b) the ``episodes'' loop in each $k$-th training ``epoch'', with the ``policy'' and ``state-value'' neural networks $\Pi_{\theta_{k}}$ and $V^{\Pi}_{\theta'_{k}}$.}
\label{fig:RL_scheme}
\end{center}
\end{figure}

One might indeed reformulate the QAOA minimization as an optimal control process \cite{Dalessandro2007} in which one acts sequentially
on the system in order to maximize a final reward. 
This reformulation seems particularly suited for Reinforcement Learning (RL)~\cite{Sutton_2ed, kober2013, mnih2013, silver2017}.
As schematically represented in Fig.~\ref{fig:RL_scheme}(a), at each discrete time step $t$ an ``agent'' is given some information, 
typically through measuring some observables $O_{t-1}$ on the state $S_{t-1}=|\psi_{t-1}\rangle$ of the system on which it acts (the ``environment''). 
The agent then performs an action $a_t$ --- here choosing the appropriate $(\gamma_{t}, \beta_{t})$ and applying the corresponding unitaries to the state 
--- obtaining a new state $S_{t}=|\psi_{t}\rangle$ and receiving a ``reward'' $r_{t}$, measuring the quality of the variational state constructed. 

Several questions come to mind, which have not been addressed in the recent literature on RL applied to quantum 
problems~\cite{Bukov_PRX2018,Fosel_PRX2018,Moritz_arxive2018, Khairy_arxiv2019, Riu_arxiv2019, yao2020policy}: 
{\bf i)} is such RL-assisted QAOA able to ``learn'' {\em optimal} schedules? 
{\bf ii)} Are the schedules found {\em smooth} in $t$? 
{\bf iii)} How to dwell with the fact that getting information from $|\psi_t\rangle$ involves quantum measurements which {\em destroy} the state?
{\bf iv)} Are the strategies learned easily {\em transferable} to larger systems?

In this Letter we show, on the paradigmatic example of the transverse field Ising chain, that optimal strategies --- well known in that case, see Ref.~\cite{Mbeng_arx19} ---
can be effectively learned with a simple Proximal Policy Optimization (PPO) algorithm~\cite{Schulman_PPO} employing very small neural networks (NN). 
We show that RL automatically learns {\em smooth} control parameters, hence realizing an optimal controlled digitized-QA algorithm \cite{Mbeng_arx19,Mbeng_arXiv2019b}.
By working with disordered quantum Ising chains we show that strategies ``learned'' on small samples can be successfully transferred to larger systems, hence
alleviating the ``measurement problem'': one can learn a strategy on a small problem which can be simulated on a computer, and implement it on a larger experimental
setup~\cite{Farhi_arXiv2019}. 

{\em RL-assisted QAOA --- }
To test our scheme, we apply it to the transverse field Ising model (TFIM) in one dimension, where detailed QAOA results are already known~\cite{Mbeng_arx19}.
Specifically, we define the target Hamiltonian $\Ham_{\target}=\Ho_z + h \Ho_x$ with 
\begin{equation}
\Ho_z = - \sum_{j=1}^N J_j \PauliSigma^z_j \PauliSigma^z_{j+1} \;, \hspace{5mm} 
\Ho_x = -\sum_j \PauliSigma^x_j \;.
\end{equation}
We start considering the uniform TFIM, where  $J_j =J$. 
The model has a paramagnetic ($h > J$) and a ferromagnetic ($h<J$) phase, separated by a $2^{\rm nd}$-order transition at $h=J$. 
%
%
The performance of QAOA on the uniform TFIM chain has been studied in detail in Refs.~\cite{Wang_PRA2018,Mbeng_arx19}.
Given a set of QAOA parameters $\left(\bgamma,\bbeta \right)$, we gauge the quality of the resulting state from the residual energy density 
\begin{equation} \label{eqn:eres_def}
\epsilon_\Ptrot^{\rm res}\left(\bgamma,\bbeta \right) = \frac{E_\Ptrot \left(\bgamma,\bbeta \right) - E_{\min} } {E_{\max} - E_{\min}} \;,
\end{equation}
where $E_\Ptrot \left(\bgamma,\bbeta \right) = \langle \psi_{\Ptrot}(\bgamma,\bbeta) | \Ham_{\target} | \psi_{\Ptrot}(\bgamma,\bbeta) \rangle$ 
is the variational energy, and $E_{\max}$ and $E_{\min}$ are the highest and lowest eigenvalues of the target Hamiltonian. 
Specifically,  the results presented below will concern targeting the classical state for $h=0$, although the approach can be easily extended to the case with $h>0$.
At $h=0$ the residual energy is bounded by the inequality~\cite{Mbeng_arx19}
\begin{equation}\label{eq:TFIM_eres}
\epsilon_\Ptrot^{\rm res}\left(\bgamma,\bbeta \right) \geq \left\{
\begin{array}{ll}
\frac{1}{2\Ptrot+2}	& \hspace{2mm} \mbox{if} \hspace{2mm} 2\Ptrot < N \vspace{2mm}\\
0              			& \hspace{2mm} \mbox{if} \hspace{2mm} 2\Ptrot \geq N 
  \end{array}
\right. \;,
\end{equation}
which becomes an equality if and only if $(\bgamma, \bbeta)$ are optimal QAOA parameters.  

The key ingredients of the RL-assisted algorithm, as schematized in Fig.~\ref{fig:RL_scheme}, are as follows. 
\begin{description} 
\item[State)] The {\em state} $S_t$ at time step $t=1,\dots,\Ptrot$ is encoded by the wave-function $\ket{\psi_t}$, defined iteratively as 
$\ket{\psi_t}  = \nep^{-i \beta_t \Hx} \nep^{-i \gamma_t \Hz} \ket{\psi_{t-1}}$, with 
$\ket{\psi_0} = |+\rangle= \frac{1}{\sqrt{2^N}}\bigotimes_i \left(\ket{\!\uparrow}_i + \ket{\!\downarrow}_i \right)$.
%
The agent has partial information through a number of {\em observables} $O_{t-1}$ measured on $\ket{\psi_{t-1}}$. Our choice (with $t-1\to t$) is 
\begin{equation} \label{eqn:Ot}
O_{t} = \big\{ \langle \psi_{t} | \PauliSigma^z_j \PauliSigma^z_{j+1} | \psi_{t} \rangle, \langle \psi_{t} | \PauliSigma^x_j | \psi_{t} \rangle \big\}  \;,
\end{equation}
where a single value of $j$ is enough when translational invariance is respected. 

\item[Action)] The action $a_{t}$ at time $t$ corresponds to choosing $(\gamma_{t},\beta_{t})$. 
The conditional probability of $a_t$ given the observables $O_{t-1}$ --- called ``policy'' in RL --- 
is denoted by $\Pi_{\theta}(a_t | O_{t-1})$, where $\theta$ are the parameters of a Neural Network (NN) encoding.
Our policy is stochastic, to help exploration: $\Pi_\theta(a | O)$ is chosen as a Gaussian distribution, whose mean and standard deviation are computed by the NN. 
From this, $a_t=(\gamma_{t},\beta_{t})$ is extracted. 

\item[Reward)] A reward $r_{t}$ is calculated at time $t$. 
In our present implementation, $r_{t=1,\dots,\Ptrot-1}=0$ and only $r_{\Ptrot}>0$. 
The final reward $r_{\Ptrot} = R(E_{\Ptrot})$ is associated to minimizing the final expectation value 
$E_{\Ptrot}= \langle \psi_{\Ptrot} | \Ham_{\target} | \psi_{\Ptrot} \rangle$.
%
%
Here $R(E_{\Ptrot})$ is monotonically increasing when $E_{\Ptrot}$ decreases.
Specifically, we take $R(E_{\Ptrot})=-E_{\Ptrot}$, but different non-linear choices have been tested. 

\item[Training)] 
The training process consists of a number $N_{\rm epo}$ of ``epochs'', as sketched in Fig.~\ref{fig:RL_scheme}(b).  
During each epoch the RL agent explores, with a {\em fixed} policy, 
the state-action trajectories for a certain number $N_{\rm epi}$ of ``episodes'', each episode involving $\Ptrot$ steps $t=1,\dots,\Ptrot$.
At the end of each epoch the policy is updated to favor trajectories with higher reward.
The particular RL algorithm we used is the Proximal Policy Optimization (PPO) algorithm~\cite{Schulman_PPO}, from the OpenAI SpinningUp library
\cite{SpinningUp2018}. 
PPO is an actor-critic algorithm where two independent NNs are used to parameterize the policy $\Pi_\theta(a_t | O_{t-1})$ and the state-value 
function~\cite{Sutton_2ed} $V_{\theta'}^\Pi(O_t)$. In our current implementation, $V_{\theta'}^\Pi(O_t)=\mathbb{E}^{\Pi}[r_{\Ptrot}]$ gives the expected 
reward that the system in a state with observables $O_t$ gets as it evolves with the policy $\Pi$. 
$V_{\theta'}^\Pi(O_t)$ is used to calculate the updates after each epoch~\cite{SpinningUp2018}.
In our numerical simulations, we used NNs with two fully-connected hidden layers of $32, 16$ neurons, and linear-rectification (ReLu) activation function. 
\end{description}

{\em Results ---}
In the RL training, the system is initially prepared in the state $\ket{\psi_0}=|+\rangle$, while the NNs for the policy and the state-value function are both 
initialized with random parameters.
The agent is then trained for $N_{\rm epo}=1024$ epochs, each comprising $N_{\rm epi}=100$ episodes of $\Ptrot$ steps each. 
After training, we test the RL algorithm with $\sim 50$ runs. 
\begin{figure}[ht]
\begin{center}
\includegraphics[scale=0.5]{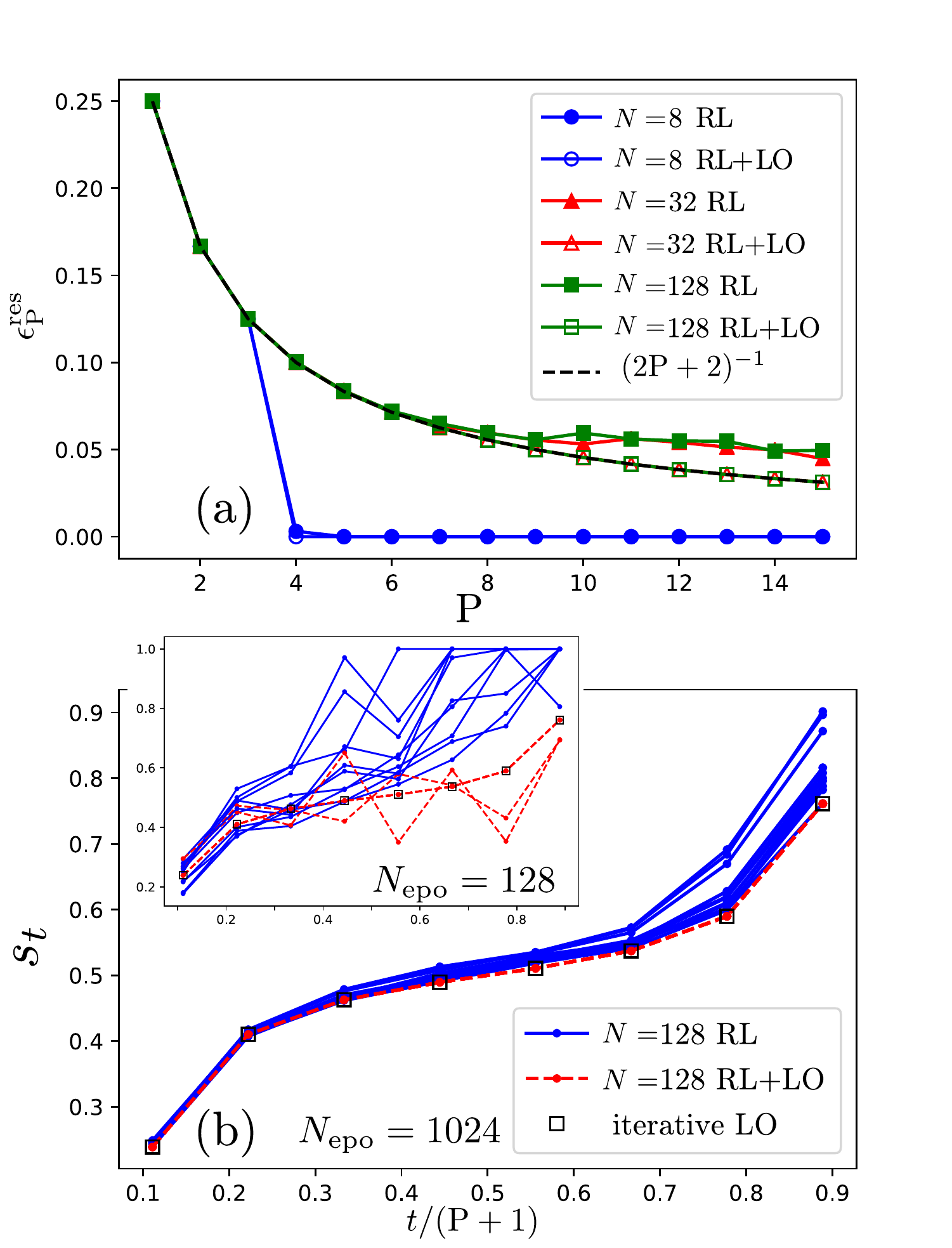}
\end{center}
\caption{(a) Residual energy density $\epsilon_\Ptrot^{\rm res}$, Eq.~\eqref{eqn:eres_def}, vs $\Ptrot$. 
Full symbols: results from RL only; empty symbols: a local optimization (LO) supplements the RL actions (RL+LO); 
data are averaged over 50 test runs.
The black dashed line is the lower bound of Eq.~\eqref{eq:TFIM_eres}. 
(b) The schedule $s_t=\gamma_t/(\gamma_t+\beta_t)$. 
Full blue lines denote $s_t$ learned after $N_{\rm epo}=1024$ epochs on a chain of $N=128$ sites; 
Dashed red lines, the RL+LO results; 
Black empty squares, the iterative LO smooth solution~\cite{Mbeng_arx19}. 
The RL actions are in the basin of the same optimal minimum.
Inset: same data for $N_{\rm epo}=128$ training epochs, where not all the LO optimized actions sets fall onto the iterative LO solution.}
\label{fig:TFIM_eres}
\end{figure}

Fig.~\ref{fig:TFIM_eres}(a) shows the results obtained by the RL-trained policy. 
For $\Ptrot \leq 6$, the trained RL agent finds optimal QAOA parameters, saturating the bound for $\epsilon_\Ptrot^{\rm res}$ in Eq.\eqref{eq:TFIM_eres}. 
In particular, for small system sizes $N$, when $\Ptrot > N/2$, the agent finds the exact target ground state, and $\epsilon_\Ptrot^{\rm res}=0$.
%
For longer episodes ($\Ptrot>6$), the residual energy deviates from the lower bound due to two factors:  
{\em i)} the longer the episode, the more difficult it is to learn the policy, as a larger number of training epochs are necessary to reach convergence;
{\em ii)} since we are using a stochastic policy, the error due to the finite width of the action distributions is accumulated during an episode, 
leading to larger relative errors for longer trajectories.
To cure this fact, we adopted the following strategy: we supplement the RL-trained policy with a final {\em local optimization} (LO) of the parameters $(\bgamma,\bbeta)$,
employing the Broyden-Fletcher-Goldfard-Shanno (BFGS) algorithm~\cite{Nocedal_book2006}.
This last step is computationally cheap, since the RL training brings the agent already close to a local minimum, provided $N_{\rm epo}$ is large enough.
The residual energy data obtained in this way, denoted by RL+LO in Fig.~\ref{fig:TFIM_eres}(a), falls on top of the optimal curve
$\epsilon_{\Ptrot}^{\rm res} = \frac{1}{2\Ptrot +2}$.

%

To visualize the action choices, we translate $\gamma_t$ and $\beta_t$ into the corresponding interpolation parameter $s_t$ which
a Trotter-digitised QA/AQC would show, which for $h=0$ is given by: \cite{Mbeng_arx19}
\begin{equation} \label{eqn:st}
s_t = \frac{\gamma_t}{\gamma_t+\beta_t} \;. 
\end{equation}
Fig.~\ref{fig:TFIM_eres}(b) shows the interpolation parameter $s_t$ during an episode $t=1,\dots,\Ptrot$, for a chain of $N=128$ spins and $\Ptrot = 8$.
Different curves are obtained by repeating a test run of the same stochastic policy, trained for $N_{\rm epo}=1024$ epochs.
The parameters obtained through the RL policy are smooth, and different tests result in similar s-shaped profiles for $s_t$.
When a final local minimization is added, the curves for $s_t$ coalesce and coincides with the smooth optimal schedule obtained in Ref.~\cite{Mbeng_arx19} through an independent iterative local optimization strategy.
When the training is at an early stage, i.e. the number of epochs is small, see inset of Fig.~\ref{fig:TFIM_eres}(b), the profiles $s_t$ are more irregular and do not fall 
all in the same smooth minimum upon performing the LO (see the three dashed red lines).

Next, we turn to the random TFIM case. Here, for each chain length $N$ we fix a given disorder instance $\lbrace J_j \rbrace_{j=1,\dots, N}$ 
with $J_j \in [0,1]$, both for the training and the test of the RL policy. 
Since translational invariance is now lost, one would naively imagine that the relevant observables $O_t$ in Eq.~\eqref{eqn:Ot} would involve a list of
$2N$ measurements. However, our experience has taught us that we can efficiently go on with a reduced list comprising only the two Hamiltonian terms, 
$O_t = \big\{ \langle \psi_t | \Ho_z | \psi_t \rangle , \langle \psi_t | \Ho_x | \psi_t \rangle \big\}$, hence chain-averaged quantities.
All the parameters involved in training the NNs 
are fixed as in the uniform TFIM case. 

\begin{figure}
\begin{center}
\includegraphics[scale=0.5]{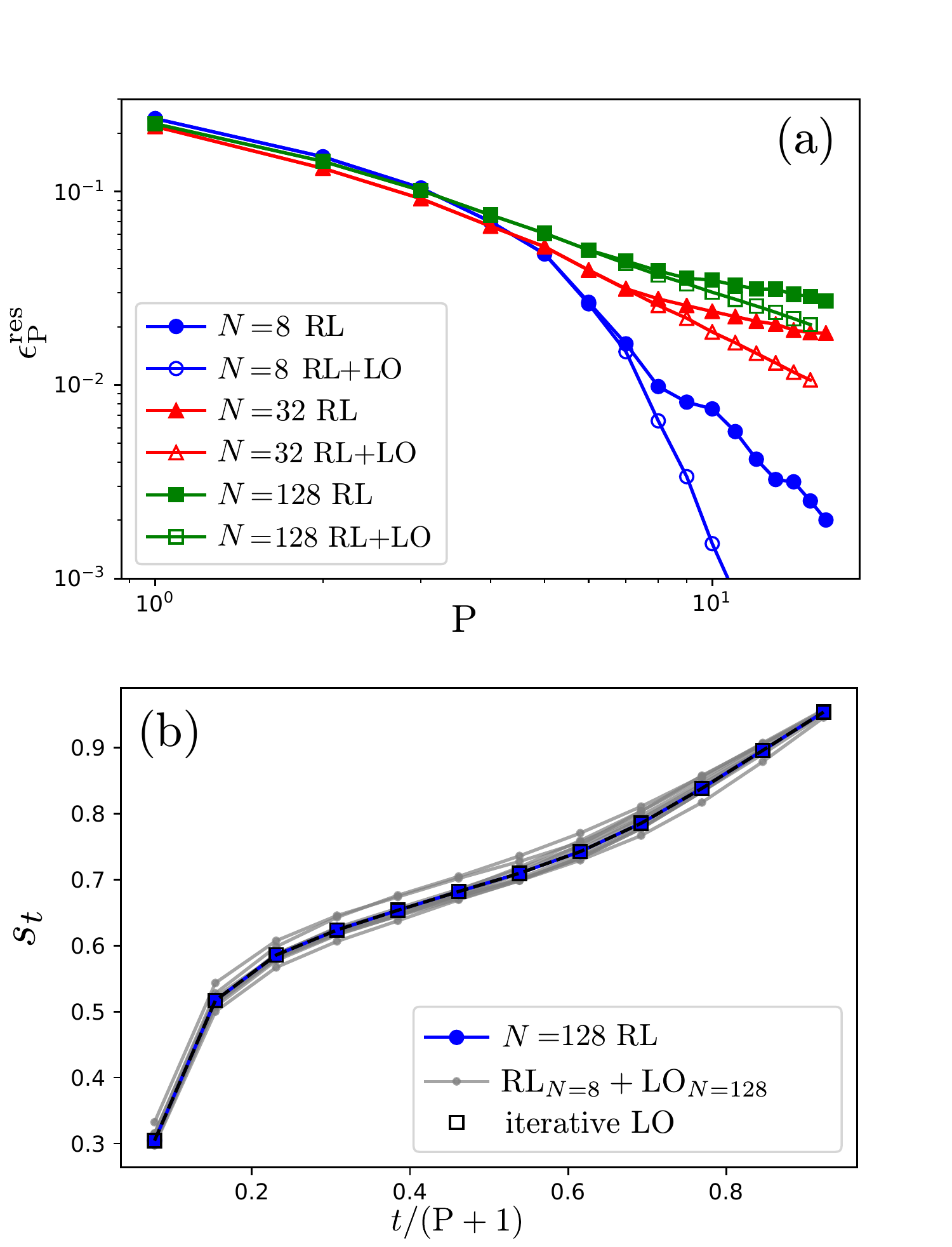}
\end{center}
\caption{(a) Residual energy, Eq.~\eqref{eqn:eres_def}, vs $\Ptrot$ for a single instance of the random TFIM: 
comparison between bare RL and RL followed by local optimization (LO) results (RL+LO). 
(b) The optimized $s_t$ obtained with different procedures.
Empty squares: the iterative LO process of Ref.~\cite{Mbeng_arx19}; 
Blue circles: RL+LO performed directly on a $N=128$ chain;
Gray lines: $\mathrm{RL}_{\scriptscriptstyle N=8}\!+\!\mathrm{LO}_{\scriptscriptstyle N=128}$,  
i.e., training of a $N=8$ chain used as {\em Ansatz} for LO of the $N=128$ chain.
}
\label{fig:rTFIM_eres}
\end{figure}

Fig.~\ref{fig:rTFIM_eres}(a) shows the residual energy $\epsilon_{\Ptrot}^{\rm res}$ vs $\Ptrot$ obtained from the bare RL (full symbols) and from RL followed by a local optimization
(RL+LO, empty symbols). 
The local optimization significantly improves the quality for large $\Ptrot \geq 10$. 
A detailed study of the behaviour of $\epsilon_{\Ptrot}^{\rm res}$ for large $\Ptrot$ and a comparison with the results obtained~\cite{Caneva_PRB07} by a 
linear-QA/AQC scheme, with $s(t)=t/\tau$, is left to a future study.  
%

Fig.~\ref{fig:rTFIM_eres}(b) shows the optimal parameter $s_t=\gamma_t/(\gamma_t+\beta_t)$ found by the RL+LO method, compared to the $s_t$ 
constructed with the iterative optimization strategy described in  Ref.~\cite{Mbeng_arx19}: the agreement between the two is remarkable, showing that the
RL-assisted QAOA effectively ``learns'' smooth action trajectories. 

The most remarkable fact, however, is shown by the series of grey lines present in Fig.~\ref{fig:rTFIM_eres}(b).
These are obtained by training the RL agent on a much smaller instance with $N=8$ sites, and transferring the RL-policy to the larger (and different) disorder instance
with $N=128$, followed by local optimizations of the learned parameters.  
These results show a large transferability of the RL policies, which we have verified to hold even in the absence of the final LO. 
This suggests the following way-out from the ``measurement problem'' involved in the construction of the state observables $O_t$. 
Indeed, in an experimental implementation of RL-assisted QAOA, the RL agent could observe a small system, efficiently simulated on a classical 
hardware, and then use the learned actions to evolve the larger experimental system. 
This reduces drastically the number of measurements to be performed and allows to test RL-assisted QAOA on physical quantum platforms.

{\em Conclusions ---}  
In this Letter we have shown that the optimal QAOA strategies well known for the TFIM~\cite{Mbeng_arx19}
can be effectively learned with a simple PPO-algorithm~\cite{Schulman_PPO} employing rather small NNs. 
The observables measured on a state, referring to the two competing terms in the Hamiltonian and providing information to the ``agent'', 
seem to be effective in the learning process.
We have shown that RL learns {\em smooth} control parameters, hence realizing an RL-assisted feedback Quantum Control for the schedule $s(t)$
of a digitized QA/AQC algorithm \cite{Mbeng_arx19}, in absence of any spectral information. 
By working with disordered quantum Ising chains we showed that strategies ``learned'' on small samples can be successfully transferred to larger systems, hence
alleviating the ``measurement problem'': one can learn a strategy on a small problem simulated on a computer, and implement it on a larger experimental
setup. 

A discussion of previous RL-work on quantum systems is here appropriate. RL as a tool for quantum control and quantum-error-correction has been investigated in Refs.~\cite{Bukov_PRX2018,Fosel_PRX2018}. 
%
%
Regarding applications to QAOA, Refs.~\cite{Moritz_arxive2018, Khairy_arxiv2019, yao2020policy} have all formulated RL strategies to learn optimal variational 
parameters $(\bgamma,\bbeta)$. 
While sharing similar RL tools, their approach is markedly different from ours: they identify the RL ``state'' with the whole set of QAOA parameters. 
The agent has no access to the internal quantum state, and no information on the evolution process can be exploited in the optimization. 
In this way, the issue of measuring the intermediate quantum state is bypassed. 
This choice, however, reduces RL to a heuristic optimization which forfeits one of the most relevant feature of the RL framework: 
The possibility to drive the process with a step-by-step evolution.
An alternative proposal, closer to ours in methods but tackling different physical questions, has recently appeared in Ref.~\cite{Riu_arxiv2019}. 


Concerning future developments, we mention possible improvements of the ``measurement problem''. 
One possibility is to introduce ancillary bits to provide intermediate information to the RL agent without destroying the state of the system, 
in a way similar to Ref.~\cite{Fosel_PRX2018}. 
A possible alternative is to perform weak measurements~\cite{Lucignano_SciRep2019}.
A second issue is the sensitivity to noise: preliminary results show that noise in the initial state preparation does not harm the
ability to learn the correct strategies. 
%
%
Finally, the application to other models is worth pursuing: preliminary results on the fully-connected p-spin Ising ferromagnet are encouraging. 

We thank R. Fazio for stimulating discussions. Research was partly supported by EU Horizon 2020 under ERC-ULTRADISS, Grant Agreement No. 834402.
GES acknowledges that his research has been conducted within the framework of the Trieste Institute for Theoretical Quantum Technologies (TQT).

\end{document}